\documentclass[zpreprint,zbstepj]{zeus_paper}
\usepackage[english]{babel}
\newcommand{\ZcoosysB}{%
The ZEUS coordinate system is a right-handed Cartesian system, with the $Z$
axis pointing in the proton beam direction, referred to as the ``forward
direction'', and the $X$ axis pointing left towards the centre of HERA.
The coordinate origin is at the nominal interaction point.\xspace}

\newcommand{\ZcoosysfnB}{\footnote{\ZcoosysB}}

\newcommand{\Zdetdesc}{%
A detailed description of the ZEUS detector can be found 
elsewhere~\cite{zeus:1993:bluebook}. A brief outline of the 
components that are most relevant for this analysis is given
below.\xspace}
\newcommand{\Zctddesc}[1]{%
Charged particles are tracked in the central tracking detector (CTD)~\citeCTD,
which operates in a magnetic field of $1.43\Tesla$ provided by a thin 
superconducting solenoid. The CTD consists of 72~cylindrical drift chamber 
layers, organized in nine~superlayers covering the polar-angle#1 region 
\mbox{$15^\circ<\theta<164^\circ$}. The relative transverse-momentum 
resolution for full-length
tracks is $\sigma(p_T)/p_T=0.0058p_T\oplus0.0065\oplus0.0014/p_T$,
with $p_T$ in $\Gev$. 
The tracking system was used to establish the primary and secondary vertices. 
}
\newcommand{\Zcaldesc}{%
The high-resolution uranium--scintillator calorimeter (CAL)~\citeCAL consists 
of three parts: the forward (FCAL), the barrel (BCAL) and the rear (RCAL)
calorimeters. Each part is subdivided transversely into towers and
longitudinally into one electromagnetic section (EMC) and either one (in RCAL)
or two (in BCAL and FCAL) hadronic sections (HAC). The smallest subdivision of
the calorimeter is called a cell.  The CAL energy resolutions, as measured under
test-beam conditions, are $\sigma(E)/E=0.18/\sqrt{E}$ for electrons and
$\sigma(E)/E=0.35/\sqrt{E}$ for hadrons, with $E$ in $\Gev$.}

\chardef\usc=95
\chardef\til=126
\catcode`\@=11 
\DeclareRobustCommand\xdotspace{\futurelet\@let@token\@xdotspace}
\def\@xdotspace{%
  \ifx\@let@token.\else
  \ifx\@let@token\bgroup.\else
  \ifx\@let@token\egroup.\else
  \ifx\@let@token\/.\else
  \ifx\@let@token\ .\else
  \ifx\@let@token~.\else
  \ifx\@let@token!.\else
  \ifx\@let@token,.\else
  \ifx\@let@token:.\else
  \ifx\@let@token;.\else
  \ifx\@let@token?.\else
  \ifx\@let@token/.\else
  \ifx\@let@token'.\else
  \ifx\@let@token).\else
  \ifx\@let@token-.\else
  \ifx\@let@token\@xobeysp.\else
  \ifx\@let@token\space.\else
  \ifx\@let@token\@sptoken.\else
   .\space
   \fi\fi\fi\fi\fi\fi\fi\fi\fi\fi\fi\fi\fi\fi\fi\fi\fi\fi}
\catcode`\@=12 

\newcommand{\stru}[2]{%
   \relax\ifmmode\hbox{\vrule height#1 depth#2 width0pt}%
   \else\vrule height#1 depth#2 width0pt\fi}

\newcommand{\Ronum}[1]{\uppercase\expandafter{\romannumeral#1}}
\newcommand{\ronum}[1]{\expandafter{\romannumeral#1}}
\DeclareRobustCommand{\LaTeXZ}{%
  \LaTeX\kern-.05em4\kern-.1em
  {\raisebox{-0.2ex}{$\scriptstyle\text{ZEUS}$}}\xspace}

\DeclareMathAlphabet{\mathbf}{OT1}{cmr}{bx}{sl}
\newcommand{\eVdist}{\kern-0.06667em}

\newcommand{\Gev}{{\text{Ge}\eVdist\text{V\/}}}

\newcommand{\Tesla}{\,\text{T}}

\newcommand{\slashfrac}[2]{%
  \raisebox{0.5ex}{\ensuremath #1}\kern-0.12em/\kern-0.08em
  \raisebox{-.8ex}{\ensuremath #2}}

\newcommand{\sqr}[3]{%
    {\vcenter{\hrule height.#3ex\hbox{\vrule width.#2ex height#1ex
     \kern#1ex\vrule width.#3ex}\hrule height.#2ex}}}

\catcode`\@=11 
\newcommand{\parenbar}{\mathpalette\p@renb@r}
\def\p@renb@r#1#2{\vbox{%
  \ifx#1\scriptscriptstyle \dimen@.7em\dimen@ii.2em\else
  \ifx#1\scriptstyle \dimen@.8em\dimen@ii.25em\else
  \dimen@1em\dimen@ii.4em\fi\fi \offinterlineskip
  \ialign{\hfill##\hfill\cr
    \vbox{\hrule width\dimen@ii}\cr
    \noalign{\vskip-.3ex}%
    \hbox to\dimen@{$\mathchar300\hfil\mathchar301$}\cr
    \noalign{\vskip-.3ex}%
    $#1#2$\cr}}}
\catcode`\@=12 

\newcommand{\IP}{{\rm I$\kern-0.01667em$P}\xspace}

\mathchardef\qsm=63
\mathchardef\pls=43
\mathchardef\mns=512
\mathchardef\plm=518
\mathchardef\eql=61
\mathchardef\smallleft=300
\mathchardef\smallright=301
\mathchardef\les=316
\mathchardef\gre=318
\mathchardef\leq=532
\mathchardef\grq=533

\catcode`\@=11 
\newcounter{pict@width}
\newcounter{pict@height}
\newlength{\pict@scale}
\newcommand{\psfigadd}[4]{%
\setcounter{pict@width}{1*\ratio{#2+\pict@scale/2}{\pict@scale}}
\setcounter{pict@height}{1*\ratio{#3+\pict@scale/2}{\pict@scale}}
\setlength{\unitlength}{\pict@scale}
\hbox to #2{\hspace{-\fill}\begin{picture}(\thepict@width,\thepict@height)
\put(0,0){\psfig{figure=#1,width=#2,height=#3,clip=}}
\SetScale{0.283466457}
\SetWidth{1.763889}
{#4}
\end{picture}}
}
\newcounter{pict@widthfst}
\newcounter{pict@widthscd}
\newcounter{pict@widthtot}
\newcommand{\psfigaddtwo}[7]{%
\setcounter{pict@widthfst}{1*\ratio{#2+\pict@scale/2}{\pict@scale}}
\setcounter{pict@widthscd}{1*\ratio{#2+#4+\pict@scale/2}{\pict@scale}}
\setcounter{pict@widthtot}{1*\ratio{#2+#4+#6+\pict@scale/2}{\pict@scale}}
\setcounter{pict@height}{1*\ratio{#3+\pict@scale/2}{\pict@scale}}
\setlength{\unitlength}{\pict@scale}
\hbox{\hspace{-\fill}\begin{picture}(\thepict@widthtot,\thepict@height)
\put(0,0){\psfig{figure=#1,width=#2,height=#3,clip=}}
\put(\thepict@widthscd,0){\psfig{figure=#5,width=#6,height=#3,clip=}}
\SetScale{0.283466457}
\SetWidth{1.763889}
{#7}
\end{picture}}
}
\newcommand{\psfigror}[4]{%
\setcounter{pict@width}{1*\ratio{#2+\pict@scale/2}{\pict@scale}}
\setcounter{pict@height}{1*\ratio{#3+\pict@scale/2}{\pict@scale}}
\setlength{\unitlength}{\pict@scale}
\hbox{\begin{picture}(\thepict@width,\thepict@height)
\put(0,\thepict@height){\psfig{figure=#1,width=#3,height=#2,clip=,angle=270}}
\SetScale{0.283466457}
\SetWidth{1.763889}
{#4}
\end{picture}}
}
\newcommand{\psfigrol}[4]{%
\setcounter{pict@width}{1*\ratio{#2+\pict@scale/2}{\pict@scale}}
\setcounter{pict@height}{1*\ratio{#3+\pict@scale/2}{\pict@scale}}
\setlength{\unitlength}{\pict@scale}
\hbox{\begin{picture}(\thepict@width,\thepict@height)
\put(0,0){\psfig{figure=#1,width=#3,height=#2,clip=,angle=90}}
\SetScale{0.283466457}
\SetWidth{1.763889}
{#4}
\end{picture}}
}
\catcode`\@=12 
\newlength\listtextwidth

\catcode`\@=11 
\newlength{\@tabfninsert}
\newlength{\@tabfnwidth}
\newcommand{\tabfootnote}[2]{%
  \setlength{\@tabfninsert}{0.8em}
  \setlength{\@tabfnwidth}{\textwidth}
  \addtolength{\@tabfnwidth}{-\@tabfninsert}
  \addtolength{\@tabfnwidth}{-0.4em}
  \noindent\makebox[\@tabfninsert][r]{\footnotesize$^{#1}$\hfil}\hfill%
  \parbox[t]{\@tabfnwidth}{\footnotesize #2\hfill}}
\catcode`\@=12 
\def\citeCTD{{\cite{%
nim:a279:290,*npps:b32:181,*nim:a338:254%
}}\xspace}
\def\citeCAL{{\cite{%
nim:a309:77,*nim:a309:101,*nim:a321:356,*nim:a336:23%
}}\xspace}
\def\citeClose93{{\cite{%
pl:b319:291%
}}\xspace}
\def\citebwigner{{\cite{%
nc:56:269%
}}\xspace}
\def\citeWeinstein90{{\cite{%
prd:v41:2236%
}}\xspace}
\def\citeJaffe77{{\cite{%
prd:v15:267%
}}\xspace}
\def\citeAlford00{{\cite{%
np:b578:367%
}}\xspace}

\def\citedjango6{{\cite{%
cpc:81:381,*spi:www:djangoh11%
}}\xspace}
\def\citePDG{{\cite{%
epj:c15:1%
}}\xspace}

\def\citeDA{{\cite{%
proc:hera:1991:23-tmp-3c83b9d7%
}}\xspace}

\def\citePDG02{{\cite{%
pr:d66:01%
}}\xspace}

\def\citeL3{{\cite{%
pl:b501:173%
}}\xspace}

\def\citeWAf1710{{\cite{%
pl:b453:305%
}}\xspace}
\def\citeGlueball{{\cite{%
pr:d60:034509,*np:b314:347%
}}\xspace}

\def\citeBES96{{\cite{%
prl:v77:3959%
}}\xspace}
\def\citeJB{{\cite{%
proc:epfacility:1979:391%
}}\xspace}
\def\citeDA{{\cite{%
proc:hera:1991:23%
}}\xspace}
\def\citemark3{{\cite{%
prl:v56:107%
}}\xspace}

\def\citeBreit{{\cite{%
bk:1972:1,*zerw:zfp:237%
}}\xspace}

\def\citeArmenteros{{\cite{%
phima:45:13%
}}\xspace}

\def\citeH1Kaons{{\cite{%
np:b480:3,*zfp:c76:213%
}}\xspace}
\def\citeKtclus{{\cite{%
np:b406:187%
}}\xspace}

\newcommand{\fprime}{$f_2^{\prime}(1525)$}

\begin{document}



\title{
Observation of $K_s^0K_s^0$ resonances \\
in deep inelastic scattering at HERA
}                                                       
                    
\author{ZEUS Collaboration}

\abstract{
Inclusive $K_s^0K_s^0$ production
in deep inelastic $ep$ scattering at HERA has been studied
with the ZEUS detector 
using an integrated luminosity of $120$~pb$^{-1}$. 
Two states are observed at masses of 
$1537^{+9}_{-8}$~MeV and $1726\pm 7$~MeV, 
as well as an enhancement around $1300$~MeV.
The state at $1537$~MeV is consistent with the well established
\fprime. The state at $1726$~MeV may be the glueball candidate
$f_0(1710)$. However, it's width of $38^{+20}_{-14}$~MeV
is narrower than $125\pm10$~MeV observed by previous experiments 
for the $f_0(1710)$.
}

\makezeustitle

                                                           
\def\3{\ss}                                                                                        
\pagenumbering{Roman}                                                                              
                                                   %
\begin{center}                                                                                     
{                      \Large  The ZEUS Collaboration              }                               
\end{center}                                                                                       
  S.~Chekanov,                                                                                     
  M.~Derrick,                                                                                      
  D.~Krakauer,                                                                                     
  J.H.~Loizides$^{   1}$,                                                                          
  S.~Magill,                                                                                       
  B.~Musgrave,                                                                                     
  J.~Repond,                                                                                       
  R.~Yoshida\\                                                                                     
 {\it Argonne National Laboratory, Argonne, Illinois 60439-4815}~$^{n}$                            
\par \filbreak                                                                                     
  M.C.K.~Mattingly \\                                                                              
 {\it Andrews University, Berrien Springs, Michigan 49104-0380}                                    
\par \filbreak                                                                                     
  P.~Antonioli,                                                                                    
  G.~Bari,                                                                                         
  M.~Basile,                                                                                       
  L.~Bellagamba,                                                                                   
  D.~Boscherini,                                                                                   
  A.~Bruni,                                                                                        
  G.~Bruni,                                                                                        
  G.~Cara~Romeo,                                                                                   
  L.~Cifarelli,                                                                                    
  F.~Cindolo,                                                                                      
  A.~Contin,                                                                                       
  M.~Corradi,                                                                                      
  S.~De~Pasquale,                                                                                  
  P.~Giusti,                                                                                       
  G.~Iacobucci,                                                                                    
  A.~Margotti,                                                                                     
  A.~Montanari,                                                                                    
  R.~Nania,                                                                                        
  F.~Palmonari,                                                                                    
  A.~Pesci,                                                                                        
  G.~Sartorelli,                                                                                   
  A.~Zichichi  \\                                                                                  
  {\it University and INFN Bologna, Bologna, Italy}~$^{e}$
% ...
\newpage                                                                                           
$^{\    1}$ also affiliated with University College London \\                                      
% ...
                                                           %
                                                           %
\newpage   
                                                           %
                                                           %
\begin{tabular}[h]{rp{14cm}}                                                                       
$^{a}$ &  supported by the Natural Sciences and Engineering Research                               
          Council of Canada (NSERC) \\                                                             
$^{b}$ &  supported by the German Federal Ministry for Education and                               
          Research (BMBF), under contract numbers HZ1GUA 2, HZ1GUB 0, HZ1PDA 5, HZ1VFA 5\\         
$^{c}$ &  supported by the MINERVA Gesellschaft f\"ur Forschung GmbH, the                          
          Israel Science Foundation, the U.S.-Israel Binational Science                            
          Foundation and the Benozyio Center                                                       
          for High Energy Physics\\                                                                
$^{d}$ &  supported by the German-Israeli Foundation and the Israel Science                        
          Foundation\\                                                                             
$^{e}$ &  supported by the Italian National Institute for Nuclear Physics (INFN) \\                
$^{f}$ &  supported by the Japanese Ministry of Education, Culture,                                
          Sports, Science and Technology (MEXT) and its grants for                                 
          Scientific Research\\                                                                    
$^{g}$ &  supported by the Korean Ministry of Education and Korea Science                          
          and Engineering Foundation\\                                                             
$^{h}$ &  supported by the Netherlands Foundation for Research on Matter (FOM)\\                   
$^{i}$ &  supported by the Polish State Committee for Scientific Research,                         
          grant no. 620/E-77/SPB/DESY/P-03/DZ 117/2003-2005\\                                      
$^{j}$ &  partially supported by the German Federal Ministry for Education                         
          and Research (BMBF)\\                                                                    
$^{k}$ &  supported by the Fund for Fundamental Research of Russian Ministry                       
          for Science and Edu\-cation and by the German Federal Ministry for                       
          Education and Research (BMBF)\\                                                          
$^{l}$ &  supported by the Spanish Ministry of Education and Science                               
          through funds provided by CICYT\\                                                        
$^{m}$ &  supported by the Particle Physics and Astronomy Research Council, UK\\                   
$^{n}$ &  supported by the US Department of Energy\\                                               
$^{o}$ &  supported by the US National Science Foundation\\                                        
$^{p}$ &  supported by the Polish State Committee for Scientific Research,                         
          grant no. 112/E-356/SPUB/DESY/P-03/DZ 116/2003-2005,2 P03B 13922\\                       
$^{q}$ &  supported by the Polish State Committee for Scientific Research,                         
          grant no. 115/E-343/SPUB-M/DESY/P-03/DZ 121/2001-2002, 2 P03B 07022\\                    
\end{tabular}                                                                                      


\newpage
\pagenumbering{arabic} 
\pagestyle{plain}
\section{Introduction}
\label{sec-int}

The $K_s^0K_s^0$ system is expected to couple to scalar
and tensor glueballs. This has motivated intense 
experimental and theoretical study during the past few years
\cite{rmp:v71:1411,hep-ex/0101031}.
Lattice QCD calculations \citeGlueball
predict the existence of a scalar 
glueball with a mass of $1730\pm 100$ MeV and 
a tensor glueball at $2400\pm 120$ MeV.
The scalar glueball can mix with $q\overline{q}$ states with $I=0$ from 
the scalar meson nonet, leading to three $J^{PC}=0^{++}$ states, whereas
only two can fit into the nonet.
Experimentally, four states with $J^{PC}=0^{++}$ and $I=0$ have been 
established \citePDG02: $f_0(980)$, $f_0(1370)$, $f_0(1500)$ and $f_0(1710)$.

The state most frequently considered to be a glueball candidate 
is $f_0(1710)$ \citePDG02, but its gluon content has not 
yet been established. 
This state was first observed in radiative $J/\psi$ 
decays \citeBES96 and
its angular momentum $J=0$ was established by the WA102 experiment 
using a partial-wave analysis in the $K^+K^-$ and $K_s^0K_s^0$ 
final states\citeWAf1710.
A recent publication from L3 \citeL3
reports the observation of two states in $\gamma\gamma$ collisions
above $1500$ MeV, the well established $f_2^\prime(1525)$ \citePDG02
and a broad resonance at 1760 MeV. 
It is not clear if this state is the $f_0(1710)$.
Observation of $f_0(1710)$ in $\gamma\gamma$ 
collisions would indicate a large quark content.

The $ep$ collisions at HERA provide an opportunity to study resonance
production in a new environment. The production of $K_s^0$ 
has been studied previously at HERA \cite{zfp:c68:29,*epj:c2:77,np:b480:3,*zfp:c76:213}.
In this paper, the first observation of resonances in the $K_s^0K_s^0$ final 
state in inclusive $ep$ deep inelastic scattering (DIS) is reported.

\section{Experimental set-up}
\label{sec-exp}

The data were collected by the ZEUS detector at HERA during the 96-00
running periods. In 96-97, HERA collided \mbox{27.5 GeV} positrons
with \mbox{820 GeV} protons. In 98-00, the proton energy was 
920 GeV and both positrons and electrons were collided with protons. The
measurements for $e^+p$ ($e^-p$) interactions\footnote{Hereafter, both
$e^+$ and $e^-$ are referred to as electrons, unless explicitly stated
otherwise.} are based on an integrated luminosity of $104
\,\mathrm{pb^{-1}}$ ($17 \,\mathrm{pb^{-1}}$).

\Zdetdesc

\Zctddesc \ZcoosysfnB

\Zcaldesc

The energy of the scattered electron was corrected
for energy loss in the material between the interaction
point and the calorimeter using a small-angle rear tracking detector
(SRTD)\cite{nim:a401:63, epj:c21:443} and a presampler (PRES) 
\cite{nim:a382:419, epj:c21:443}.

\section{Kinematic reconstruction and event selection}
\label{kinematic}

The inclusive neutral current DIS process 
$e(k)+p(P) \rightarrow e(k^\prime)+X$
can be 
described in terms of the following variables:
the negative of the invariant-mass 
squared of the exchanged virtual photon,
$Q^2~=~-q^2~=~-$ $(k-k^\prime)^2$; the fraction of the lepton 
energy transferred 
to the proton in the proton rest frame,
$y=(q\cdot P)/(k\cdot P)$;
and the Bjorken scaling variable, 
$x=Q^2/(2P\cdot q)$. 
 
A three-level trigger system was used to select events online 
\cite{zeus:1993:bluebook}.
The inclusive DIS selection was defined by requiring an
electron found in the CAL. 
In certain run periods, corresponding to 83\% of the total luminosity, 
the inclusive selection was not available for low $Q^2$ (\mbox{$Q^2<20$} 
GeV$^2$) events. For these periods, an additional selection with a 
requirement 
of at least one forward jet identified with the $k_t$ algorithm 
\citeKtclus and having transverse energy 
$E_T>3$ GeV ($E_T>4$ GeV in the 2000 running period), 
pseudorapidity $0~<~\eta_{\rm jet}~<~3$ 
($1.5~<~\eta_{\rm jet}~<~3.5$ in the 1996-1997 running period)
was used.

The DIS offline event selection was based on the following requirements:
\begin{itemize}
\item
a primary vertex position, determined from the tracks fitted to the 
vertex, in the range $|~Z_{\rm vertex}~|~<~50~$cm, to reduce the
background events from non-{\it ep} interactions;
\item
$E_{e}~\ge~8.5$ GeV, where $E_{e}$ is the energy of the 
scattered electron reconstructed in the calorimeter; 
\item
$42~<~\delta~<60$ GeV, where $\delta~=~\sum E_i(1~-~{\rm cos}\theta_i)$, 
$E_i$ is the energy of the $i^{th}$ calorimeter cell and $\theta_i$ is 
its polar angle as viewed from the 
primary vertex. The sum runs over all cells. 
This cut further reduces the background from
photoproduction and events with large QED initial-state radiation;
\item
$y_{\rm e}~\le~0.95$, to remove events with misidentified scattered DIS 
electrons; $y_{\rm e}$ is the value of {\it y} 
reconstructed using the scattered electron measurements;
\item
$y_{\rm JB}~\ge~0.04$, to remove events with low 
hadronic activities; 
$y_{\rm JB}$ is the value of {\it y} reconstructed using the Jaquet-Blondel 
method \citeJB ;
\item
the position of the scattered lepton candidate in the RCAL was required to 
be outside a box of $\pm 14$ cm in {\it X} and {\it Y}, which corresponds 
approximately to \mbox{$\theta_{\rm ele} \approx 176^o$}, where 
$\theta_{\rm ele}$ is the polar angle of the scattered electron;
\item
a maximum of 40 tracks per event. This cut reduces
the background from false $K^0_s$ pair candidates, 
removing only 7\% of the DIS events.
\end{itemize}

The non-ep and photoproduction background in the selected
sample was negligible.
The data were not corrected
for the biases introduced by the trigger requirements and selection cuts.

\section{Selection of $K_s^0$-pair candidates}
\label{kkselection}
The $K_s^0$ meson candidates were reconstructed using tracks with at
least 38 hits in the CTD and pseudorapidity within $\pm 1.75$.
The pseudorapidity cut and the required minimum number of hits ensured a 
good momentum resolution and a minimum transverse momentum of 0.1 GeV. 
In each event,
oppositely charged track pairs assigned to a secondary vertex were
combined to form $K_s^0$ candidates. Both tracks were assigned the mass 
of a charged pion and the invariant-mass $M(\pi^+\pi^-)$ calculated.
The secondary vertex resolution for these events, 
estimated using MC studies, is 2 mm in X and Y, and 4 mm in 
Z.

Additional requirements were applied to the selected 
$K_s^0$ candidates:

\begin{itemize}

\item
$p_T(K_s^0)~>~200$~MeV, for each $K_s^0$ candidate;

\item
$2~<~d~<~30~$cm, where $d$ is the decay length
of the $K_s^0$ candidate;
\item
$d_{XY}~<~4~$mm and $d_{Z}~<~5.5~$mm, where $d_{XY}$ and $d_{Z}$ are,
respectively, the 
projections on the XY plane and Z axis of the vector
defined by the primary interaction point and the point
of closest approach of the $K_s^0$ candidate;
\item
$\theta_{XY}
~<~0.12$, where $\theta_{XY}
$ is 
the (collinearity) angle between the candidate $K_s^0$ 
momentum vector and the vector defined by the interaction point and the 
$K_s^0$ decay
vertex in the $XY$ plane;
\item
$p_T^A>110$ MeV, for each $K_s^0$ candidate, 
where the Armenteros-Podolanski variable $p_T^A$ is the 
projection of the candidate pion momentum onto
a plane perpendicular to the $K_s^0$ candidate line of flight
\citeArmenteros.

\end{itemize}
 
The cuts on the decay length, distance of closest approach and collinearity 
angle significantly reduce the non-$K_s^0$ background  
as determined by Monte Carlo (MC) simulations. 
After the  $p_T^A$ cut, backgrounds from $\Lambda$, $\overline\Lambda$ 
and photon conversions are negligible.
Only events with at least two selected $K_s^0$ candidates were kept for
further analysis.
 
\begin{figure}[p]
\vfill
\begin{center}
\epsfig{file=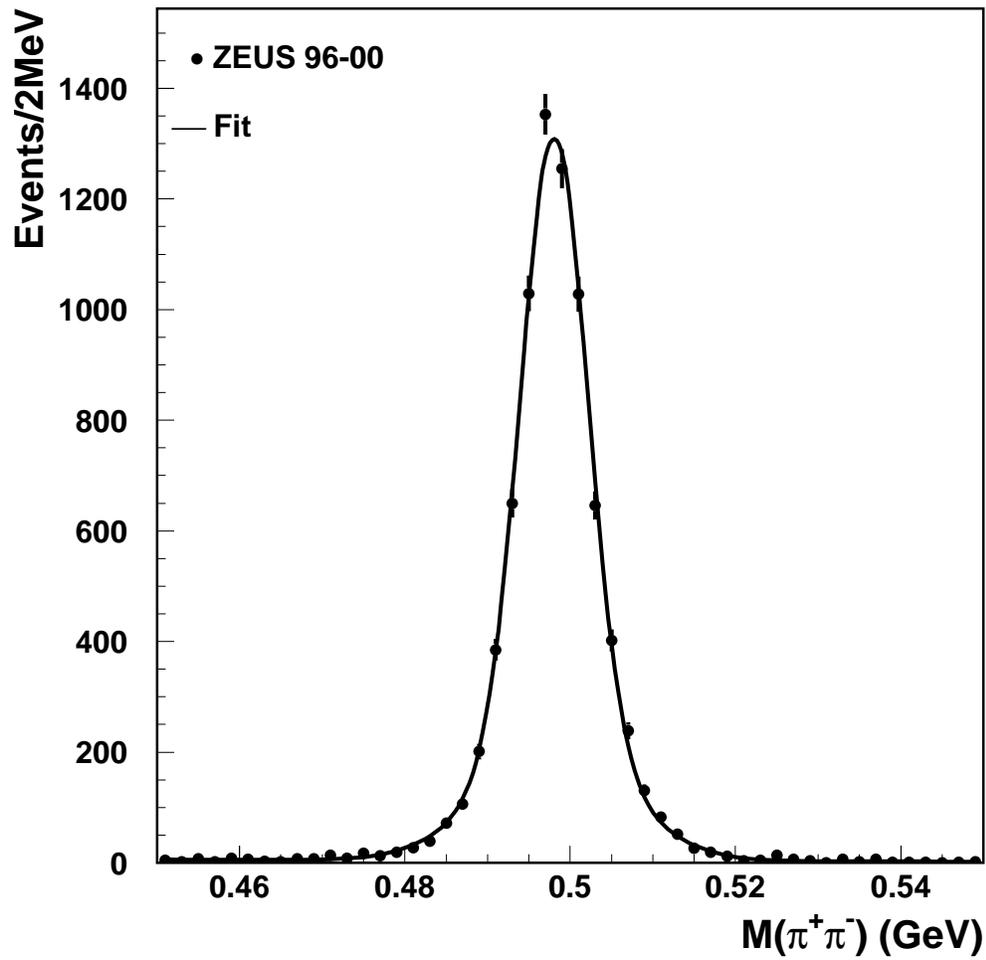,width=0.9\textwidth}
\end{center}
 \caption{The distribution of $\pi^+\pi^-$ invariant mass 
for events with at least two $K_s^0$ candidates 
passing all selection cuts. The solid line shows the result of a
fit using one linear and two Gaussian functions.
\label{fig:Ksmass} }
\vfill
\end{figure}

Figure \ref{fig:Ksmass} shows the invariant-mass distribution for 
$K_s^0$ candidates in the range $0.45<M(\pi^+\pi^-)<0.55$ GeV after
the $K_s^0$-pair candidate selection.
The distribution was fitted using one linear and two Gaussian functions. 
The linear function fits the background, one of the
Gaussians fits the peak region in the central $\pi^+\pi^-$ invariant
mass distribution, and the other Gaussian improves the fit at the tails. 
The two Gaussians were constrained to have the same mean value.
A mass of 498 MeV and standard deviation width of 4 MeV were obtained from 
the fit with the central Gaussian, and a standard deviation width of 8.5 MeV
was obtained from the fit with the other Gaussian.
The normalization factor between the narrower and broader Gaussians
is approximately 5.5. 
The invariant-mass width is dominated by the momentum 
resolution of tracks reconstructed with the CTD.
Only $K_s^0$ candidates in the region of $\pm 10$ MeV around the 
fitted central mass 
were used to reconstruct the $K_s^0K_s^0$ invariant-mass.
High-statistics samples of Monte Carlo events, generated
without resonances, were used to confirm that the event-selection
criteria did not produce artificial peaks in the $K_s^0K_s^0$
invariant-mass spectrum.

\begin{figure}[p]
\vfill
\begin{center}
\epsfig{file=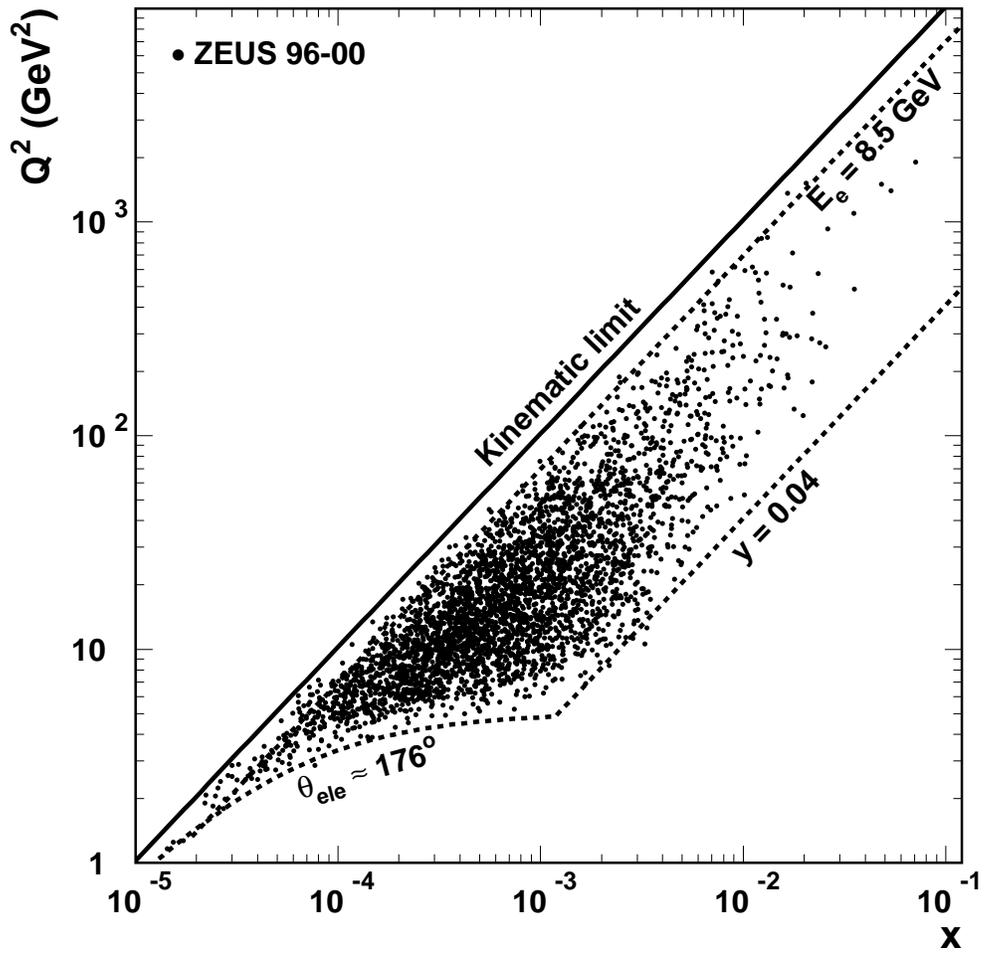,width=0.9\textwidth}
\end{center}
 \caption{The distribution in $x$ and $Q^2$ of events passing all selection cuts.
The dashed lines delineate approximately the 
kinematic region selected. The solid
line indicates the kinematic limit for HERA running with 920 GeV protons.
\label{fig:q2vsxbj} }
\vfill
\end{figure}

Figure \ref{fig:q2vsxbj} shows the distribution in $x$ and $Q^2$ 
of selected events containing at least one 
pair of $K_s^0$ candidates. The kinematic variables were reconstructed 
using the double angle method \citeDA.
The virtual-photon proton 
centre-of-mass energy was in the range \mbox{$50<W<250$ GeV}.

\section{Results}
\label{sec-kk}
%
%

The $K_s^0K_s^0$ spectrum may have a strong enhancement near the
$K_s^0K_s^0$ threshold due to the $f_0(980)$/$a_0(980)$
state \cite{prd:v32:189,pl:b489:24,*hep-ph/0202157}.
Since the high $K_s^0K_s^0$ mass 
is the region of interest for this analysis, the complication 
due to the threshold region is avoided by imposing the cut 
\mbox{$cos\theta_{K_s^0K_s^0}<0.92$}, where $\theta_{K_s^0K_s^0}$ 
is the opening angle between the two $K_s^0$ candidates in the 
laboratory frame. 

\begin{figure}[p]
\vfill
\begin{center}
\epsfig{file=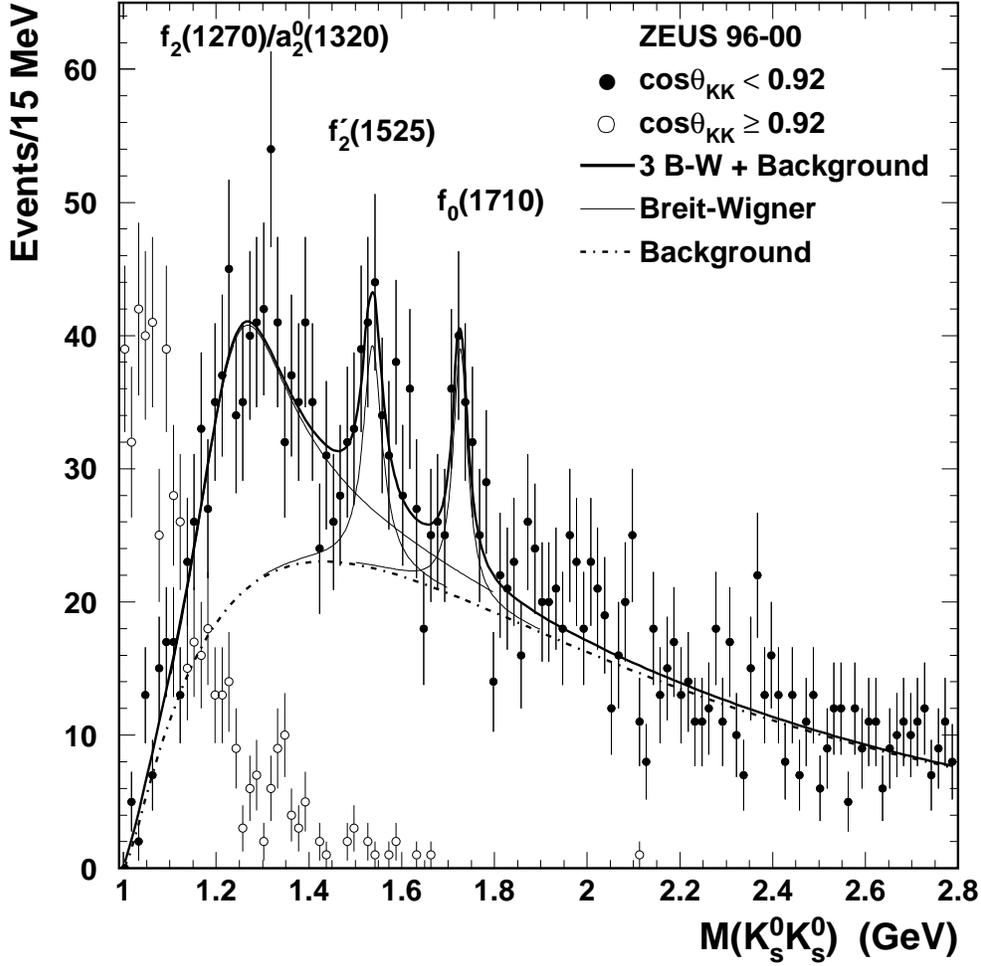,width=0.9\textwidth}
\end{center}
 \caption{ The $K_s^0K_s^0$ invariant-mass spectrum for $K_s^0$ pair candidates with
$cos\theta_{K_s^0K_s^0}<0.92$ (filled circles). 
The thick solid line is the result 
of a fit using three Breit-Wigners (thin solid lines) and a background
function (dotted-dashed line). 
The $K_s^0$ pair candidates that fail the $cos\theta_{K_s^0K_s^0}<0.92$
cut are also shown (open circles).
\label{fig:KKmass} }
\vfill
\end{figure}

After applying all selections, 2553 $K_s^0$-pair candidates were
found in the range
$0.995<M(K_s^0K_s^0)<2.795$ GeV, where
$M(K_s^0K_s^0)$ was calculated using the $K_s^0$ mass of $497.672~$MeV
\citePDG02. The momentum resolution of the CTD leads to an average
$M(K_s^0K_s^0)$ resolution which ranges from $7$ MeV in the 
$1300$ MeV mass region to 10 MeV in the $1700$ MeV region.
Figure \ref{fig:KKmass} shows the measured $K_s^0K_s^0$ invariant-mass 
spectrum.
Two clear peaks are 
seen, one around 1500 MeV and the other around 1700 MeV,
along with an enhancement around 1300 MeV.
The data for \mbox{$\cos{\theta_{K_s^0K_s^0}}>0.92$} are also
shown. The mass scale uncertainties in the region of interest,
arising from uncertainties in the magnetic field, is at the
per mille level.

The distribution of Fig.~\ref{fig:KKmass} was fitted 
using three modified relativistic Breit-Wigner (MRBW) 
distributions 
and a background function $U(M)$;

\begin{equation}
F(M) = \sum_{i=1}^{3}
(\frac{m_{*,i}\Gamma_{d,i}}{(m_{*,i}^2-M^2)^2+m_{*,i}^2\Gamma_{d,i}^2})
+ U(M)~,
\end{equation}

where $\Gamma_{d,i}$ is the effective resonance width, 
which takes into account spin and large width effects \citebwigner,
$m_{*,i}$ is the resonance mass, and $M$ is $K^0_sK^0_s$ invariant 
mass. The background function is

\begin{equation}
U(M)~=~A\cdot (M-2m_{K_s^0})^{B}\cdot e^{-C \sqrt{M-2m_{K_s^0}}}~,
\label{bgndfunc}
\end{equation}

where $A$, $B$ and $C$ are free parameters and 
$m_{K_s^0}$ is the $K_s^0$ mass defined by Hagiwara et al \citePDG02. 
Monte Carlo studies showed that effects of the track-momentum resolution
on the mass reconstruction were small compared to the measured widths 
of the states. Therefore the resolution effects were ignored in the fit.

Below $1500$ MeV, a region strongly affected by the 
$\cos{\theta_{K_s^0K_s^0}}$ cut, a peak is seen around 
$1300$ MeV where a contribution from $f_2(1270)/a_2^0(1320)$ is expected.
This mass region was fitted with a single Breit-Wigner.

Above $1500$ MeV, the lower-mass state has a fitted mass of 
$1537^{+9}_{-8}$ MeV and a width of $50^{+34}_{-22}$ MeV, in good 
agreement with the well established $f_2^\prime(1525)$.
The higher-mass state has a fitted mass of $1726 \pm 7$ MeV and
a width of $38^{+20}_{-14}$ MeV.
The widths reported here were stable, within statistical errors,
to a wide variation of fitting methods including those using 30 MeV 
bins rather than the default 15 MeV bins.
The width is 
narrower than the PDG value
($125 \pm10$ MeV) \citePDG02 reported for $f_0(1710)$, but when 
it is fixed to this value, 
the fit is still acceptable. For the purposes of the following 
discussion, this state is referred to as $f_0(1710)$.

The masses, widths and number of events from the fit 
with statistical errors are given in the top row of Table~\ref{table1}.
It should be noted that
there are correlations between the obtained parameters. In particular the
width and the number of events for each state are highly correlated.
The mass spectrum is 
also consistent with the background function at higher masses, but masses 
above 2795 MeV were not included in the fit due to limited statistics.
The sensitivity of the data to the widths of the resonances was studied. 
Several fits were performed fixing the width of the states 
$f_2^\prime(1525)$ and $f_0(1710)$ to their PDG values; the results 
are shown in rows 2 to 4 of Table \ref{table1}.

In the literature, there are several states reported in the mass region near 
2000 MeV \citePDG02, namely $f_2(1950)$ and $f_0(2020)$ which need confirmation, 
and $f_2(2010)$, $a_4(2040)$ and $f_4(2050)$ which have been confirmed. 
While no visible structure in this mass region exist in the data, these states may
affect the background shape.
The effect of the inclusion of a state in this region 
was examined, with the result shown in row 5 of Table \ref{table1}.
The description of the data between 1800 and 2000 MeV is improved with
respect to the other fits. However, there is no improvement in the overall
$\chi^2/dof$. 

The $K_s^0K_s^0$ spectrum after all cuts was also fitted
with the background function only. 
The fit can be rejected with a 99.4\% confidence level using a 
simple $\chi^2$ test over the mass region 1000 to 2800 MeV.

It was found that 93\% of the $K_s^0$-pair candidates selected within the 
detector and trigger acceptance are in the target region of the Breit 
frame \citeBreit, the hemisphere containing the proton remnant. Of 
the $K_s^0$-pair candidates in the target region, 78\% are in the 
region $x_{p}=2p_B/Q > 1$, 
where $p_B$ is the absolute momentum of the $K_s^0K_s^0$ in the Breit frame.
High $x_{p}$ corresponds to production of the $K_s^0$-pair in a region
where sizeable initial state gluon radiation may be expected. This is in
contrast to the situation at $e^+e^-$ colliders where the particles 
entering the hard scattering are colourless.  

\section{Conclusions}
\label{sec-con}

The first observation in $ep$ deep inelastic scattering of a state 
at $1537$ MeV, consistent with 
$f^\prime_2(1525)$, and another at 1726 MeV, close to $f_0(1710)$, 
is reported. 
There is also an enhancement near 1300 MeV which may arise from the 
production of $f_2(1270)$ and/or $a_2^0(1320)$ states.
The width of the state at $1537$ MeV
is consistent with the PDG value for the $f_2^\prime(1525)$.
The state at 1726 MeV has a mass consistent with the glueball candidate
$f_0(1710)$, and is found in a gluon-rich region of phase space.
However, it's width of $38^{+20}_{-14}$ MeV is narrower than the PDG value 
of 125$\pm$10 MeV for the $f_0(1710)$.

%
\section{Acknowledgements}
We thank the DESY directorate for their strong support and 
encouragement. The special efforts of the HERA machine group 
in the collection of the data used in this paper are 
gratefully acknowledged. We are grateful for the support of the
DESY computing and network services. The design, construction 
and installation of the ZEUS detector have been made 
possible by the ingenuity and effort of many people from DESY 
and home institutes who are not listed as authors.
We also thank F. Close, S. Godfrey and H. Lipkin for their 
valuable comments and advice.



\providecommand{\etal}{et al.\xspace}
\providecommand{\coll}{Collab.\xspace}
\catcode`\@=11
\def\@bibitem#1{%
\ifmc@bstsupport
  \mc@iftail{#1}%
    {;\newline\ignorespaces}%
    {\ifmc@first\else.\fi\orig@bibitem{#1}}
  \mc@firstfalse
\else
  \mc@iftail{#1}%
    {\ignorespaces}%
    {\orig@bibitem{#1}}%
\fi}%
\catcode`\@=12



\begin{sidewaystable} 
  \small
  \centering
\begin{tabular}{|c|c|c|c|c|c|c|c|c|c|c|c|c|c|} \hline
Fit
&$\chi^2/N$ 
&\multicolumn{3}{c|}{ $f_2(1270)/a_2^0(1320)$}
&\multicolumn{3}{c|}{$f_2^\prime(1525)$} 
&\multicolumn{3}{c|}{ $f_0(1710)$} 
&\multicolumn{3}{c|}{ $f_J(1980)$} 
\\ \hline 
&
& mass & width & events 
& mass & width & events 
& mass & width & events 
& mass & width & events 
\\ \hline
1 & 0.97
& $1274^{+17}_{-16}$ & $244^{+85}_{-58}$   & $414^{+184}_{-125}$ 
& $1537^{+9}_{-8}$   & $50^{+34}_{-22}$    & $84^{+41}_{-31}$ 
& $1726\pm 7$   & $38^{+20}_{-14}$    & $74^{+29}_{-23}$ 
&    &    &   
\\ \hline
2 & 0.96
& $1272\pm 16$ & $240^{+76}_{-55}$   & $420^{+167}_{-122}$ 
& $1539\pm 10$ & $76$                & $107\pm 30$ 
& $1727\pm 7$   & $39\pm 20$    & $76^{+28}_{-24}$ 
&    &    &   
\\ \hline
3 & 1.02
& $1276\pm 16$ & $258^{+80}_{-59}$   & $480^{+190}_{-141}$ 
& $1536\pm 8$   & $49^{+30}_{-21}$    & $85^{+38}_{-27}$ 
& $1726\pm 13$ & $125$               & $122\pm 40$ 
&    &    &   
\\ \hline
4 & 1.02
& $1274\pm 15$ & $251^{+72}_{-55}$   & $476^{+176}_{-131}$ 
& $1538\pm 10$ & $76$                & $108^{+31}_{-29}$  
& $1728\pm 13$ & $125$               & $120^{+41}_{-38}$ 
&       &    &   
\\ \hline
5 & 1.00
& $1283\pm 15$ & $260^{+70}_{-55}$   & $506^{+218}_{-122}$ 
& $1540^{+12}_{-10}$ & $70^{+43}_{-30}$    & $116^{+59}_{-42}$ 
& $1727\pm 7$   & $47^{+23}_{-15}$    & $91^{+34}_{-26}$ 
& $1970^{+33}_{-45}$ & $138^{+173}_{-89}$  & $74\pm 40$ 
\\ \hline
\multicolumn{14}{|c|}{Particle Data Group 2002 Values (MeV) \citePDG02 } \\ \hline
&
&\multicolumn{3}{c|}{ $1275.5 \pm1.2$, $185^{+3.4}_{-2.6}$ }
&\multicolumn{3}{c|}{ $1525 \pm5$, $76 \pm10$              } 
&\multicolumn{3}{c|}{ $1713 \pm6$, $125 \pm10$             } 
&\multicolumn{3}{c|}{                                      } 
\\  
&
&\multicolumn{3}{c|}{ $1318 \pm0.6$, $104.7 \pm1.9$  }
&\multicolumn{3}{c|}{              } 
&\multicolumn{3}{c|}{              } 
&\multicolumn{3}{c|}{              } 
\\ \hline 
\hline

\end{tabular}
  \caption{\it 
Normalized $\chi^2$, masses in MeV, widths in MeV, 
and number of events extracted from the $K_s^0K_s^0$ invariant mass
fits (errors are statistical only.) 
Widths reported without
errors were fixed to the PDG values listed in the last row.}
\label{table1}
\end{sidewaystable} 


%
\end{document}